\documentclass[aps,prb,twocolumn,amsmath,amssymb]{revtex4}
\usepackage{epsfig}
\usepackage{graphicx}

\begin{document}

\title{Granularity and vortex dynamics in LaO$_{0.92}$F$_{0.08}$FeAs as probed by harmonics of the AC magnetic susceptibility}

\author{Maria G. Adesso$^{1,2}$, Danilo Zola$^{1,2}$, Jianlin Luo$^{3}$, G. F. Chen$^{3}$, Zheng Li$^{3}$, N. L. Wang$^{3}$,
Sandro Pace$^{1,2}$, Canio Noce$^{1,2}$ and Massimiliano Polichetti$^{1,2}$}
\affiliation{$^1$Laboratorio Regionale SuperMat, CNR-INFM, Baronissi (SA), Italy \\
$^2$Dipartimento di Fisica ``E. R. Caianiello'', Universit\`a di
Salerno, I-84081 Baronissi (Salerno), Italy \\ $^3$Beijing National Laboratory for Condensed Matter Physics, Institute of Physics, Chinese Academy of Sciences, Beijing 100190, China}
\date{\today}

\begin{abstract}

Fundamental and higher harmonics of the AC magnetic susceptibility have been measured on a LaO$_{0.92}$F$_{0.08}$FeAs sample as a function of the temperature, at various amplitudes and frequencies of the AC magnetic field, with a small superimposed DC field parallel to the AC field. The granularity of the sample has been investigated and the inter-grain and intra-grain contributions have been clearly individuated looking at both the first and third harmonics. The vortex dynamics has been also analyzed, and a comparison with the magnetic behavior of both the MgB$_2$ and the cuprate superconductors has been performed. Some vortex dissipative phenomena, i.e. the thermally activated flux flow  and the flux creep, have been detected in the presented measurements, similar to what obtained on YBCO. Nevertheless, although the general behavior is similar, several differences have been also evidenced between these different classes of superconductors, mainly in the third harmonics. We infer that different vortex dynamics have to be included into the analysis of the magnetic response in this iron-based new material.

\end{abstract}

\pacs{74.70.Dd, 74.25.Ha, 74.25.Bt, 74.25.Qt}

\maketitle

\section{Introduction}

Recently, a relatively high temperature superconductor with a critical temperature transition as higher as 26~K, with an onset
temperature at 32~K, has been observed in LaO$_{1-x}$F$_x$FeAs family\cite{kamihara08}  provoking a deep impact in condensed-matter physics community. Indeed,
this new superconductor does not belong to any known
categories of high-temperature superconductors such
as copper-oxides materials,\cite{bednorz86} fullerenes,\cite{hebard91} and MgB$_2$\cite{nagamatsu01}.

The discovery of this new family of layered superconductors containing the magnetic elements like Fe or Ni\cite{watanabe07} raises questions about the relationship
between magnetism and superconductivity, the origin
of the remarkably high-$T_c$, and the chemical and structural
parameters that can be used to tune the properties. Concerning the iron-based family, the undoped LaOFeAs is a metal, or a degenerate semiconductor, at room temperature with
no indication of superconductivity, whereas substitution of oxygen with fluorine atoms gives rise to the superconductivity at the F doping of x=0.03. Further doping causes a gradual
increase of superconducting transition temperature up to the maximum of 26~K at x = 0.11.\cite{kamihara08}

LaOFeAs crystallizes in a tetragonal layered structure of
the P4/nmm symmetry\cite{zimmer95} (ZrCuSiAs-type structure\cite{johnson74}) made of alternating LaO and FeAs layers stacked along the $c$ axis.
The crystal structure is therefore rather simple with eight atoms in
two formula units in the cell and two internal parameters
for the position of the La and Fe atoms. The Fe atoms are coordinated by four As atoms in a distorted tetrahedron geometry, two different As-Fe-As bonding angles
being observed in the x-ray diffraction experiments.
Moreover, the lanthanum atoms are surrounded by four As atoms
and four O atoms in a distorted square antiprism, while the
O atoms have four La neighbors arranged in a tetrahedron.
Nevertheless, although the crystal structure is substantially different from that of cuprate superconductors, both compounds share intriguing
similarities such as (i) the two-dimensional crystal/electronic structure\cite{singh08,sefat08},
(ii) the presence of a superconducting dome in the electronic phase diagram where the $T_c$ is controlled by a systematic aliovalent ion doping into the
insulating block layers,\cite{chen08} and  (iii) a characteristic anomaly in
the transport property in the underdoped region.\cite{delacruz08}

However, many issues still remain open. Indeed, concerning the superconducting phase it is not known the symmetry of superconducting gap as well as the pairing mechanism. Moreover, the relevance of spin fluctuations above $T_c$, the role played by the magnetic rare-earth elements and the effect of electronic correlations deserve more investigation both theoretically and experimentally.

Therefore, these new superconductors could offer the possibility to better understand the mechanisms of the superconductivity both in these materials and in the cuprates. Moreover the determination of the actual value of their coherence length is still an open question and it does not exclude that they could be promising for the developments of applications.

As in all type-II superconductors, a limitation to the use of these superconducting materials is due to the dissipative phenomena associated with the vortex motion inside the sample.\cite{tinkham} Therefore, in order to improve both the fabrication process and the application perspectives of these materials, a crucial point is to investigate their flux pinning and vortex dynamics properties.  Among the experimental techniques used in the past, one of the most efficient is the AC magnetic susceptibility.
In this work we report an analysis of the vortex dynamics starting from measurements of fundamental and higher harmonics of the AC magnetic susceptibility. This technique has been often used in the past to investigate superconducting properties\cite{gomory97,ling91} and, sometimes, it has been also used to study the pinning properties\cite{digioacchino99}, the vortex dynamics\cite{qin00,polichetti00,polichetti03,adesso04,adesso03} and the external magnetic field versus temperature phase diagram\cite{adesso06}, in several type-II superconductors, both low-$T_c$\cite{adesso06} and high-$T_c$\cite{polichetti00,polichetti03,adesso04,adesso03}, as well as MgB$_2$\cite{adesso04}.

The paper is organized as follows. In next section we describe the preparation of LaO$_{0.92}$F$_{0.08}$FeAs sample (LaOFFeAs) and illustrate the experimental setup; a motivation of the relevance of the harmonics study is also presented.
Then, we have investigated the temperature dependence of the first and third harmonics of the AC magnetic susceptibility, and the Cole-Cole plots, i.e. the imaginary part of the harmonics as a function of the real part. More specifically, in the section III the granularity of the sample has been investigated and a direct comparison with a granular YBCO has been also performed. The vortex dynamics has been investigated in the sections IV and V, looking at the analysis of the amplitude (section IV) and frequency (section V) dependence of the harmonics of the AC susceptibility. Finally, section VI contains the conclusions.

\section{Sample preparation, experimental setup and AC magnetic susceptibility}
The polycrystalline samples were prepared by the solid state reaction using LaAs, Fe$_2$O$_3$, Fe and LaF$_3$ as starting materials. Differently from the synthesis method reported by Kamihara {\it et al.}\cite{kamihara08}, we use Fe$_2$O$_3$ as a source
of oxygen instead of La$_2$O$_3$ due to the high stability of lanthanum oxide. Lanthanum arsenide (LaAs) was obtained by reacting La chips and As pieces at 500 $^{\circ}$C for
12h then at 850 $^{\circ}$C for 2h. Mixtures of four components were ground thoroughly and cold-pressed into pellets. The pellets were placed into Ta crucible and sealed
in quartz tube under argon atmosphere. They were then annealed for 50h at a temperature of 1150 $^{\circ}C$. The phase purity was checked by a powder X-ray diffraction method using Cu K${\alpha}$ radiation at room temperature\cite{chen08}.

The first and third harmonics of the AC magnetic susceptibility have been measured on a LaOFFeAs bulk sample, as a function of the temperature ($T$) by using a Quantum Design PPMS-6000 equipped with a susceptometer, at various frequencies $\nu$ and amplitudes $h_{AC}$ of the AC magnetic field, in presence of a DC field (H$_{DC}$=50~Oe) parallel to the AC field. Before each measurement the residual field trapped inside the superconducting DC magnet is reduced to a value H$_{DC}<$1~Oe and the sample is warmed up to a temperature sufficiently higher than the superconducting critical temperature $T_c$. After that, it is first cooled in zero AC and DC fields down to the starting temperature (depending on the frequency and on the ac field amplitude, due to the presence of heating effect which could produce problems in the temperature stability of the sample), remaining about 20 minutes at that temperature in order to ensure the thermal homogeneity of the sample. After that, the fields are applied and all the harmonics data are acquired contemporarily and for increasing temperature, each point being recorded by using a "five points" measurement mode. With this mode, each single measurement is taken first at the bottom detection coil of the AC susceptometer insert, then at the Top detection coil, then again at the bottom detection coil, and finally two times at the center of the coil array. In this way, any spurious signal background or effect due to an eventual thermal drift or temperature instability is canceled.

In order to understand the fundamental mechanisms governing the flux lines dynamics of type-II superconductors in the mixed state, their magnetic properties have been studied both using DC and AC magnetic techniques, resulting in a number of information showing that the magnetic response of a superconductor can be both linear and nonlinear.
The linear case is characterized by the absence of harmonics higher than the first one in the AC magnetic susceptibility. In particular, it is known that real part of the first harmonic, associated to the screening properties of the sample, is proportional to the time average of the magnetic energy stored in the volume occupied by the sample, whereas the imaginary part is proportional to the energy converted into heat during one cycle of AC field.\cite{gomory97,clem88} The shape of these two components of the first harmonics as function of the temperature does not substantially change if the external measurement parameters (i.e. the frequency and the amplitude of the AC field, or the intensity of an eventual superimposed DC field) are changed: in practice, the curve of real part of susceptibility $\chi_1^{\prime}(T)$  always shows a step-like transition, corresponding to a curve characterized by a peak in the imaginary part $\chi_1^{\prime \prime}(T)$  curve, independently of the experimental parameters thar influence the width of the two curves only. This means that, using the first harmonics components only, it is not simple to detect the different vortex dynamics regimes governing the AC response of a sample corresponding to the different measurement conditions.
On the other side, when a state with irreversible magnetic properties is present in a material, its current-voltage ($I-V$) characteristic exhibits a non linear behavior. This determines the presence of nonlinearities in the magnetic response of the sample which are associated to the appearance of harmonics higher than the first one. On the contrary, the only presence of high-order harmonics does not imply an irreversibility in the magnetic properties of the sample. The physical interpretation given to each of the higher harmonics of the AC susceptibility is still under discussion, as well as the strong differences in the shape of their curves measured in different conditions still need clarifications. Nevertheless, by looking only at the third harmonics, both experimental and numerical results have shown that, for a sample with given geometry and pinning properties, the shape of the two components $\chi_3^{\prime}(T)$   and  $\chi_3^{\prime \prime}(T)$  is much influenced by the presence of particular vortex dynamics regimes.\cite{digioacchino99,adesso03}
So, the higher sensitivity of the third harmonics to the vortex dynamics, together with the opportunity to compare their simulated and measured curves, supplies a very powerful tool to extract detailed information about the dynamic regimes governing the AC magnetic response of superconducting samples.
Finally, the plot of the imaginary part as function of the real part of the harmonics (both the first and the third) produces the Cole-Cole plot, and this kind of representation gives additional information with respect to the plot of the separate components as function of the temperature. In particular, by using this approach it is possible to distinguish also among the different models describing a specific flux dynamics regime which mainly influences the magnetic response of a sample.

\section{$\chi_1$ and $\chi_3$: evidence of granularity}

In Fig. 1 a typical behavior of the real and imaginary part of both the first and third harmonics is shown, as measured at $\nu$= 541Hz, $h_{AC}$ = 10Oe. In Fig. 2, the corresponding first and third harmonics Cole-Cole plots are reported.
%
\begin{figure}[!h]
\includegraphics[width=0.42\textwidth]{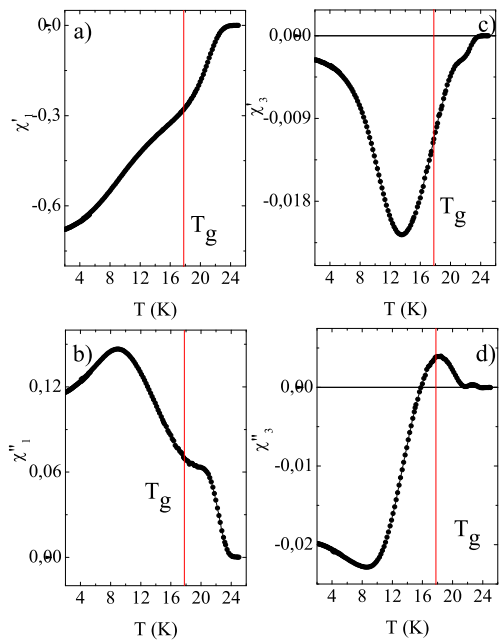}
\caption{First (a, b) and third (c, d) harmonics of the AC magnetic susceptibility as a function of the temperature, measured on LaOFFeAs, at $\nu$=541~Hz, $h_{AC}$=10~Oe. The intra-grain and the inter-grain contributions can be detected. The temperature $T_g$ at which the bulk superconductivity signal becomes predominant is also indicated in all the curves.}
\end{figure}
%

From these measurements, the granular nature of the sample can be evidenced: the contributions at high temperatures correspond to the individual grains of the material becoming superconducting (intra-granular superconductivity). In zero field cooling conditions, at these temperatures the flux lines are excluded from the inside of the grains, except a very thin surface film, so that the grains become ideally diamagnetic. At lower temperatures, corresponding to the inter-grain region, the bulk superconductivity becomes relevant. Within this temperature range, the connections between the neighbouring grains become superconducting.\cite{ozogul05} From Figs. 1 and 2 it is possible to clearly distinguish the inter-grain and the intra-grain contributions. In particular, a double step in the real part of the first harmonics and a double peak in the corresponding imaginary part are known features of a granular magnetic response.\cite{ozogul05} The temperature $T_g$ (=17.8K at this amplitude and frequency), at which the bulk superconductivity starts to prevail over the separated grain signal, has been individuated in the first harmonics curves. In the corresponding third harmonics (both in the real and in the imaginary part) it is possible to observe some more complex structures corresponding to the intra-grain (near $T_c$) and the inter-grain contributions. In particular, the real part of the third harmonics is characterized by a negative relative minimum (valley) associated to the intra-grain component and a negative absolute minimum (peak) at lower temperatures, due to the inter-grain currents. Correspondingly, the imaginary part of the third harmonics is constituted by a positive peak near $T_c$ associated to the intra-grain contribution. A further positive peak at lower temperatures is indicative of the crossover from intra to inter-grain component, its maximum corresponding to the temperature $T_g$ individuated in the first harmonics measurements. A further negative peak at lower temperatures characterizes the inter-grain contribution.

%
\begin{figure}[!t]
\includegraphics[width=0.42\textwidth]{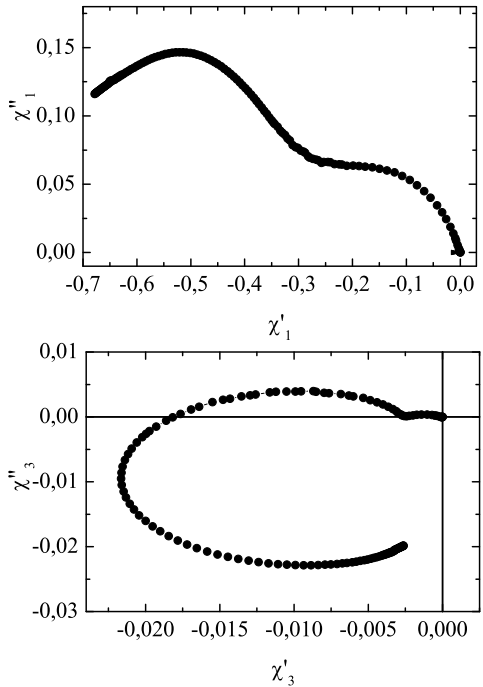}
\caption{First and third harmonics Cole-Cole plots measured on LaOFFeAs at a fixed frequency ($\nu$=541~Hz) and amplitude ($h_{AC}$=10~Oe) of the AC magnetic field. A clear distinction between the inter-grain and the intra-grain contributions can be evidenced.}
\end{figure}
%

A better way to analyze these two distinct components is by using the Cole-Cole plots, both of the first and third harmonics. As shown in Fig. 2, the first harmonics Cole-Cole plots are characterized by two separate dome-shaped curves, whereas two distinct incomplete loops are associated to the two corresponding magnetic contributions in the third harmonics Cole-Cole plots.

%
\begin{figure}[!t]
\includegraphics[width=0.52\textwidth]{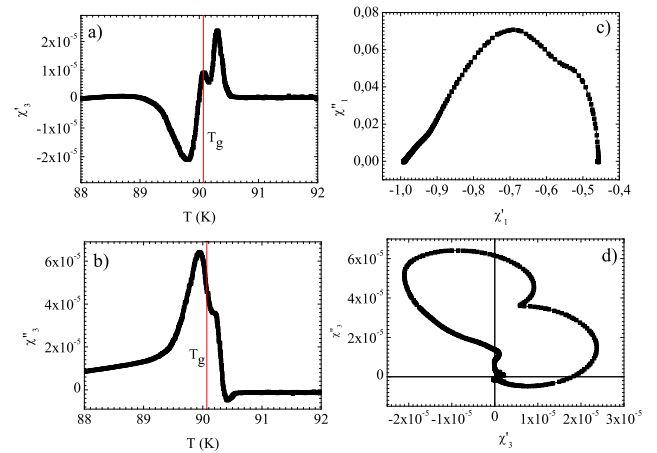}
\caption{Real (a) and imaginary (b) part of the third harmonics as a function of the temperature, as measured on a granular YBCO sample, at $\nu$=107~Hz and amplitude $h_{AC}$=16~Oe. The temperature $T_g$ has been evaluated from the temperature dependence of the first harmonics, not reported here; first (c) and third (d) harmonics Cole-Cole plots, obtained on the same sample.}
\end{figure}
%

These characteristics are typical of a type-II granular superconductor. As a comparison, in Fig. 3 some analogue measurements, performed on a commercial YBCO granular sample, are reported.
From a qualitative comparison between the measurements performed on the YBCO and the LaOFFeAs samples, we observe that the first harmonics Cole-Cole plots behave in a similar way in both the superconductors, both showing the two separate positive peaks, associated to the two contributions. Similar behaviors are also visible in the separated components of the first harmonics, not shown here for the YBCO sample.
Although the main structures associated with the granular behavior are evident in both the superconductors, a different shape in the third harmonics has been detected. From a comparison between the third harmonics of the two superconductors we infer that in the $\chi_3^{\prime}$ curves:
(a)	the intra-grain contribution in the YBCO is characterized by two structures: a positive peak near $T_c$, followed by a valley at lower temperatures, but still in the positive quadrant, whereas just a negative valley and a further decrease have been detected in the LaOFFeAs measurements;
(b)	the inter-grain contribution is characterized for both the samples by a negative peak at lower temperatures; whereas in $\chi_3^{\prime \prime}$ curves there are three main peaks in both the materials:
(i)	a small peak near $T_c$, in the intra-grain component, which is positive in the LaOFFeAs and negative in the YBCO; (ii) a positive peak in both the superconductors, which is associated with the crossover between the inter and the intra-grain contributions;
(iii) a peak in the inter-grain component, which is negative in the LaOFFeAs and positive in the YBCO sample.

Moreover, the third harmonics Cole-Cole plots are characterized, in both the type-II superconductors, by two separate loops, one associated with the intra-grain and the other one to the inter-grain contribution. In YBCO we can easily distinguish both the loops whereas in the LaOFFeAs, it is not possible to detect the full intra-grain loop. This can be ascribed to the fact that in the latter sample the full Meissner state is not achieved in our measurements. In both the samples the inter-grain loop is mainly in the left semi-plane, but in YBCO it occupies only the positive quadrant, whereas both the positive and the negative ones are occupied by the LaOFFeAs curves. Because of the not completely fulfilled intra-grain loop in the LaOFFeAs curves, the main differences between the intra-grain contribution of the two superconductors cannot be clearly evidenced.

So, although several common features between the new superconductor and a granular YBCO sample can be detected, some differences have been also evidenced and they are mainly indicative that different flux pinning properties and/or vortex dynamics are dominant in the LaOFFeAs with respect to the YBCO.

\section{Amplitude dependence of the harmonics of the AC magnetic susceptibility}

In order to analyze the vortex dynamics which main influences the magnetic response of this new superconductor, measurements at various amplitudes and frequencies of the AC magnetic field have been performed. This Section is devoted to the amplitude dependence of the harmonics, whereas next Section will contain the experimental data about the dependence on the frequency of AC magnetic field.

%
\begin{figure}[!t]
\includegraphics[width=0.42\textwidth]{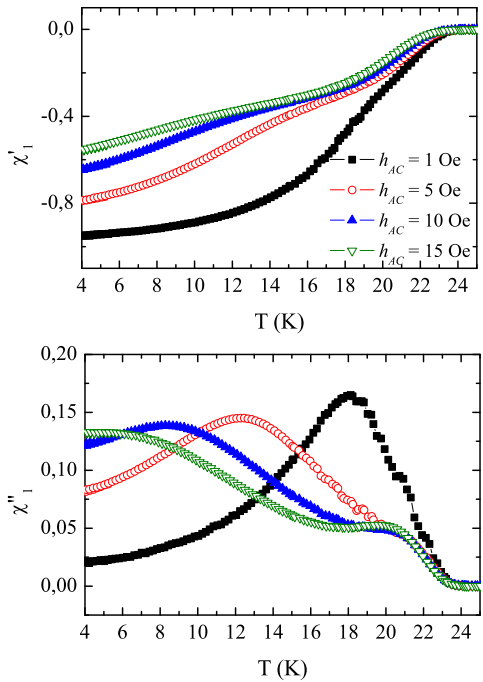}
\caption{(Color online) Temperature behavior of the real and imaginary part of the first harmonics, measured on LaOFFeAs sample at various amplitudes of the AC magnetic susceptibility, at a fixed frequency ($\nu$=107~Hz).}
\end{figure}
%

In Fig. 4 the real and imaginary part of the first harmonics as a function of the temperature is plotted at various amplitudes of the AC magnetic field, at a fixed frequency ($\nu$= 107 Hz). Similar results (not shown here) have been obtained at different frequencies. The first harmonics Cole-Cole plots are reported in Fig. 5 at various amplitudes, for given frequencies (107, 541, 1607 and 9619~Hz). In Fig. 6, the third harmonics curves are plotted, as obtained at the same external conditions of Fig. 4. A zoom of Fig. 6 about the intra-grain contribution is shown in Fig. 7. Finally, in Fig. 8 the third harmonics Cole-Cole plots are reported, at various amplitudes and frequencies of the AC field.

%
\begin{figure}[!t]
\includegraphics[width=0.42\textwidth]{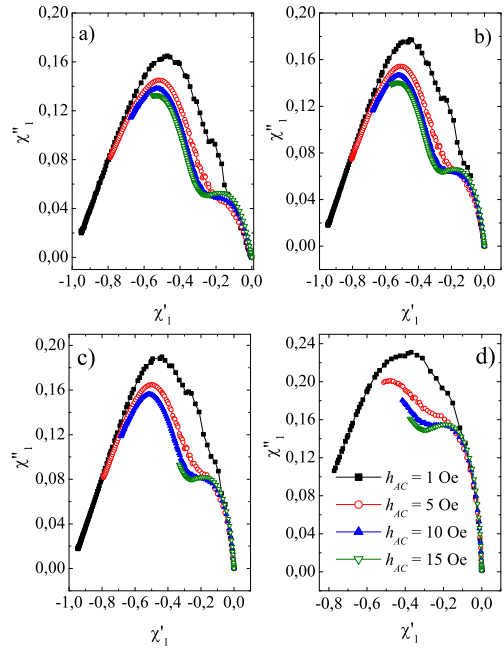}
\caption{(Color online) first harmonics Cole-Cole plots at various amplitudes of the AC magnetic field, at a fixed frequency: a)$\nu$=107~Hz, b) $\nu$=541~Hz, c) $\nu$=1607~Hz, d)   $\nu$=9619~Hz, respectively.}
\end{figure}
%

%
\begin{figure}[!b]
\includegraphics[width=0.42\textwidth]{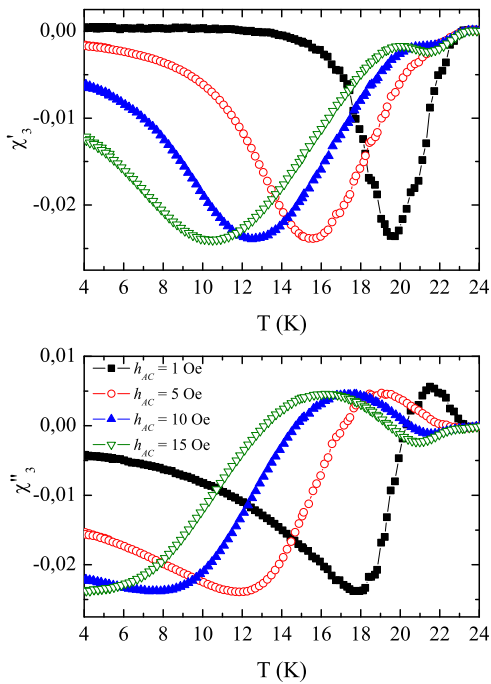}
\caption{(Color online) The real and imaginary part of the third harmonics as a function of the temperature, as measured at various $h_{AC}$  at a fixed frequency ($\nu$=107~Hz).}
\end{figure}
%

%
\begin{figure}[!t]
\includegraphics[width=0.42\textwidth]{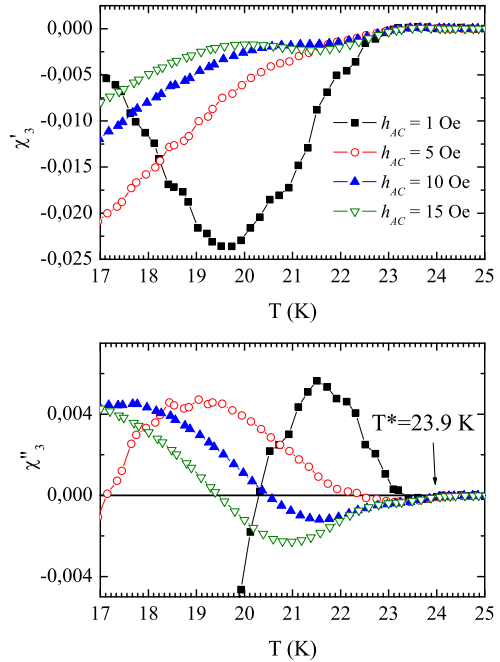}
\caption{(Color online) Enlargement of Fig. 6 in the temperature window related to the intra-grain contribution.}
\end{figure}
%

%
\begin{figure}[!t]
\includegraphics[width=0.42\textwidth]{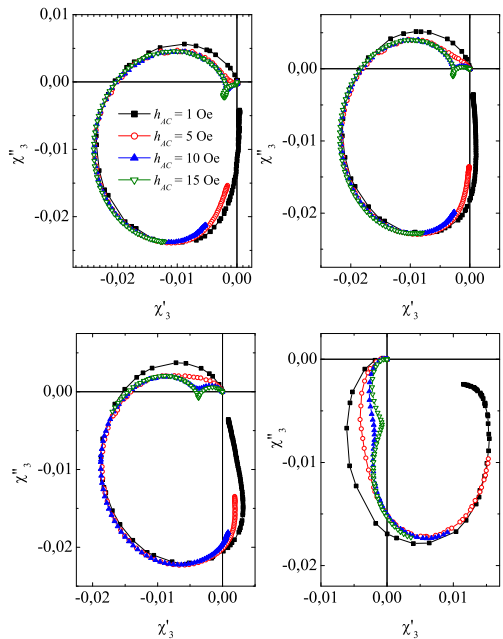}
\caption{ (Color online) Third harmonics Cole-Cole plots at various amplitudes of the AC magnetic field, at a fixed frequency: a)$\nu$=107~Hz, b) $\nu$=541~Hz, c) $\nu$=1607~Hz, d)   $\nu$=9619~Hz, respectively.}
\end{figure}
%

From Figs 4-8, we observe that, at each amplitude, it is possible to distinguish the intra-grain from the inter-grain component, with the main characteristics individuated in the previous section.
From the measurements of the first harmonics (Fig. 4) we observe that the critical temperature is independent on the amplitude of the AC magnetic field: $T_c$ = 24.6~K. On the contrary, the temperature $T_g$ decreases for increasing $h_{AC}$, with a quadratic behavior, as shown in Fig. 9, where the fit parameters are also reported.
The height of the negative bump in  $\chi_1^{\prime}$, associated with the crossover between the inter-grain and the intra-grain component, also decreases for increasing $h_{AC}$. Correspondingly, in the imaginary part of the first harmonics, the height of the peak in the inter-grain component decreases with $h_{AC}$, whereas the peak height due to the intra-grain component is characterized by a quick decrease from 1 to 5~Oe, and an almost constant value for the other amplitudes of the AC magnetic field.
In the corresponding first harmonics Cole-Cole plots (Fig. 5), the maximum in the inter-grain components decrease  with $h_{AC}$, at each frequency. A more complex behavior can be evidenced in the intra-grain components depending on the frequency.

These last characteristics furnish some indications for the analysis of the experimental curves. In the Bean model\cite{bean62} contrarily to the singular components of the harmonics, both the first and third harmonics are independent of the amplitude of the AC magnetic field, at each pinning model\cite{tesi}. Therefore, the observed amplitude dependence in the measured Cole-Cole plots indicates that vortex dissipative phenomena are occurring. In order to investigate the amplitude dependence of the harmonics of the AC magnetic susceptibility in type-II superconductors, some numerical simulations\cite{digioacchino99}$^-$\cite{adesso03} have been previously performed, by integrating the non linear diffusion equation of the magnetic field inside the sample (supposing that the sample is an infinite slab), with different $I-V$ characteristics, also by including different  pinning\cite{digioacchino99,polichetti03} and flux creep\cite{adesso04,adesso03} models. In all the simulations, a common behavior was obtained: the first harmonics Cole-Cole plots were characterized by a dome shaped curve, with a maximum decreasing with $h_{AC}$.\cite{adesso04} The data about this new superconductor in the inter-grain component are in agreement with the previously obtained simulated curves.\cite{digioacchino99}-\cite{adesso03} Nevertheless, this behavior of the inter-grain contribution with $h_{AC}$ is opposite to the experimental results obtained on different type-II superconductors, both high $T_c$ (YBCO\cite{adesso03} and BSCCO\cite{ozogul05}) and MgB$_2$\cite{adesso04}: the first harmonics Cole-Cole plots in these last materials are characterized by a maximum growing with the amplitude of the AC field. In order to reproduce the experimental results obtained on YBCO and MgB$_2$ a further dependence on $h_{AC}$  was included into the pinning potential\cite{adesso06} whereas the present measurements show that it is not necessary to include it to analyze the inter-grain behavior obtained on the LaOFFeAs sample. On the other side, the AC field intra-grain response much resembles what already found on the previously analyzed materials, thus suggesting that the pinning properties of the LaOFFeAs grains can have strong analogies with the same properties of MgB$_2$ and high-$T_c$ materials.   Nevertheless, the analysis of the first harmonics are not enough to identify the vortex dynamics which may influence the magnetic response of the sample, being this behavior common to all the numerical results obtained with different I-V curves, by using the standard pinning and creep models. Therefore, it is necessary to analyze the third harmonics curves.

A different onset temperature has been identified between the first and the third harmonics and both the temperatures $T_c$ =24.6~K and T$^{\star}$=23.9~K are independent of the amplitude of the AC magnetic field. This temperature difference between T$^{\star}$ and $T_c$ is indicative that a dissipative linear phenomenon is occurring in the temperature range [T$^{\star}$, $T_c$], supposing that the harmonics higher than the third ones are negligible with respect to the third harmonics. Both the flux flow (FF) and the thermally
activated flux flow (TAFF)  are not linear dissipative phenomena in absence of a DC field. Nevertheless, with a DC magnetic field higher than the AC one the field dependence of the FF and of the TAFF resistance disappears, generating a linear diffusion process of the magnetic field, so that there is no higher harmonics contribution. Some numerical results previously obtained\cite{digioacchino00} show that, in presence of a DC magnetic field, a characteristic temperature exists, above which the harmonics behavior is governed by the linear TAFF resistance. This temperature is nearly coincident with the onset temperature of the third harmonics. Therefore, a difference $\Delta T_{on}$ between T$^{\star}$ and $T_c$ represents the experimental evidence of the presence of linear TAFF phenomena in the system under analysis. A similar result was also obtained for YBCO bulk samples.\cite{polichetti03} 
From an analysis of the third harmonics curves, we observe that both in their real and imaginary part (Fig.6), the negative peaks in the inter-grain region become wider for increasing $h_{AC}$, but their heights (in absolute value) are almost constant. This AC field independence of the peak heights is even more clear if we observe the inter-grain loops in the third harmonics Cole-Cole plots, as shown in Fig. 8.

The presence of the inter-grain negative peak in the real part of the third harmonics, at temperatures lower than the peak in the imaginary part of the first harmonics, indicates the existence of a pronounced flux creep governing the bulk magnetic response of our sample, similar to what happens in measurements previously performed on YBCO samples.\cite{polichetti03}
The behavior of the intra-grain component with $h_{AC}$ (Fig. 7) shows that the real part of the third harmonics maintains its general shape, both the relative and the absolute minima moving towards lower temperatures for increasing amplitudes. These behaviors with $h_{AC}$ indicate that (at least in the used AC field range) no modification of the vortex dynamics regime  is produced. In the imaginary part $\chi_3^{\prime \prime}$ , the positive peak tends to become wider and to move towards lower temperatures, whereas a negative peak near $T_c$ tends to become more evident at higher $h_{AC}$, in agreement with the YBCO curves.

%
\begin{figure}
\includegraphics[width=0.42\textwidth]{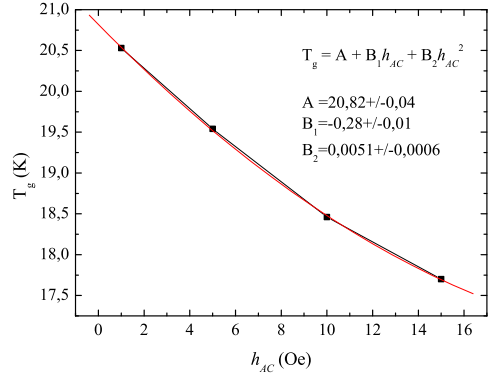}
\caption{behavior of $T_g$ as a function of $h_{AC}$, showing a quadratic decrease law. The corresponding grain critical temperature $T_c$ is independent of the amplitude of the AC field.}
\end{figure}
%

\section{Frequency dependence of the harmonics of the AC magnetic susceptibility}
In Fig. 10 both the real and the imaginary part of the first harmonics as a function of the temperature are shown, at various frequencies for a fixed amplitude ($h_{AC}$ =1~Oe) of the AC magnetic field. From Fig 10 we observe that the curves are depending on the frequency just in the temperature range 9~K-24~K; in particular, the $T_c$ is still 24.6~K, independently of the frequency. Also the temperature $T_g$ (although not easy to individuate) is independent of the frequency within the experimental errors ($T_g$=20.5~K), whereas its amplitude dependence has been reported in the previous section. Nevertheless, both  in the inter-grain and in the intra-grain component, a similar behavior with the frequency can be deduced. In particular both the peaks in the imaginary part of the first harmonics, associated to the inter-grain and the intra-grain contributions respectively, appear to shift towards higher temperatures with a growing height for increasing frequencies. This common behavior in the first harmonics inter-grain and the intra-grain components, with the frequency of the AC magnetic field, is representative that a vortex dynamical phenomenon, with a similar frequency dependence, is occurring both inside and outside the grains.

%
\begin{figure}[!t]
\includegraphics[width=0.42\textwidth]{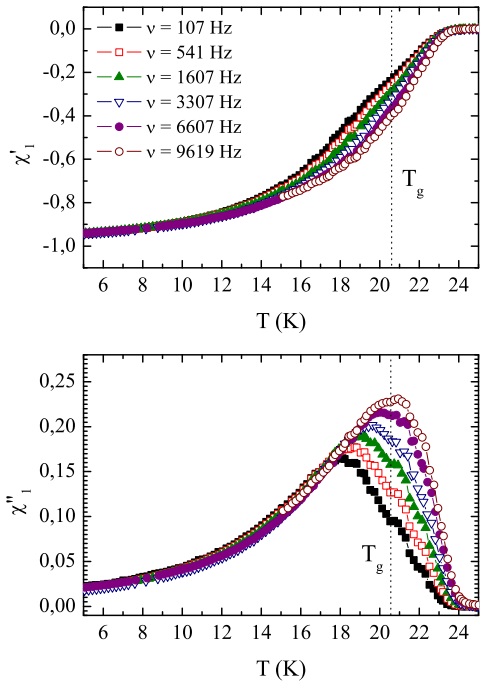}
\caption{ (Color online) Temperature behavior of the first harmonics of the AC magnetic susceptibility for a fixed amplitude ($h_{AC}$=1~Oe) at various frequencies of the AC magnetic field. The dotted line identifies the temperature $T_g$ in all the curves. }
\end{figure}
%

%
\begin{figure}[!t]
\includegraphics[width=0.42\textwidth]{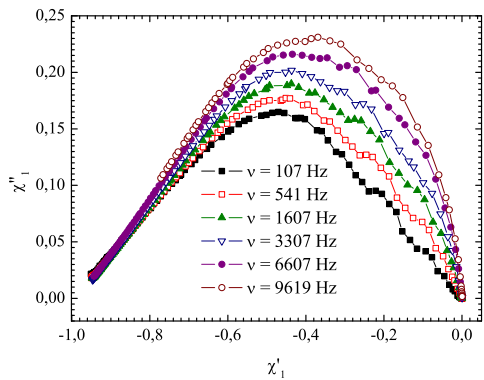}
\caption{ (Color online) First harmonics Cole-Cole plots, as measured at various frequencies of the AC magnetic field at $h_{AC}$=1~Oe. }
\end{figure}
%

This is also confirmed by the first harmonics Cole-Cole plots curves (shown in Fig. 11), where the inter-grain and the intra-grain contributions can be separately observed: in both the contributions, the peak height of the dome-shaped curves grows for increasing frequencies.

A growing height with the frequency both in the peak of the imaginary part of the first harmonics and in the maximum of the first harmonics Cole-Cole plots are both indicative of a vortex glass phase, characterized by a collective flux creep\cite{adesso08}, being this behavior opposite to that in the curves simulated by using the Kim-Anderson flux creep model.\cite{adesso08} Nevertheless, for a clear identification of the vortex glass phase, it is also necessary the analysis of  the third harmonics curves, which in our case results a quite complex task. In Fig. 12, the real and the imaginary part of the third harmonics are shown at various frequencies. A zoom of the Fig. 12 on the intra-grain component is reported in Fig. 13. The corresponding third harmonics Cole-Cole plots are reported in Fig. 14.

The real part of the third harmonics is characterized by two peaks, corresponding to the inter-grain and the intra-grain contribution. The inter-grain part of the curves is a negative peak at the lowest frequencies. For increasing frequencies its height (in the absolute value) decreases and a positive peak appears at frequencies higher than 1600 Hz. Nevertheless, in the intra-grain real part of the third harmonics (zoomed in Fig. 13) there are not shape changes: it is characterized by a negative peak with a minimum height decreasing for increasing frequencies.
%
\begin{figure}[!t]
\includegraphics[width=0.42\textwidth]{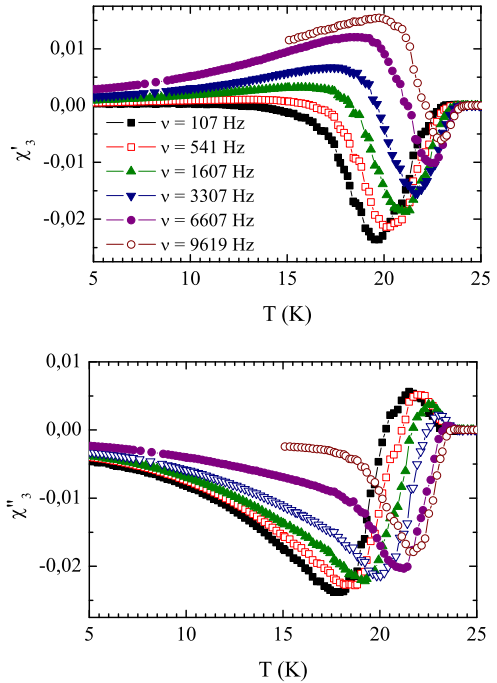}
\caption{ (Color online) Temperature behavior of the real and imaginary part of the third harmonics measured at various frequencies of the AC field at fixed $h_{AC}$=1~Oe.}
\end{figure}
%

%
\begin{figure}[!t]
\includegraphics[width=0.42\textwidth]{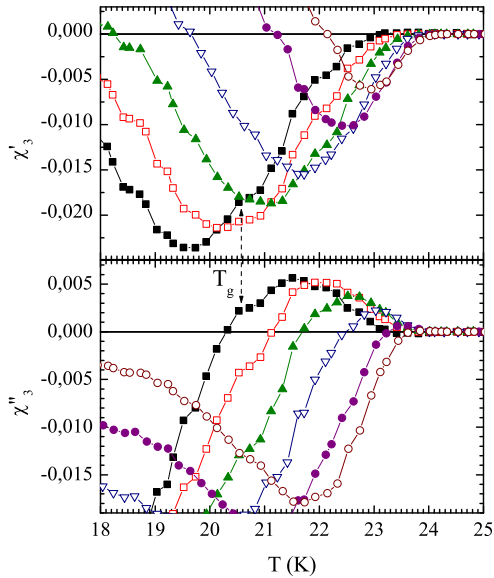}
\caption{ (Color online) A zoom of Fig. 12 in a window where the intra-grain behavior can be easily seen. The arrow across the two plots indicates the temperature $T_g$, which separates the inter-grain from the intra-grain contribution. }
\end{figure}
%

On the contrary, in the imaginary part of the third harmonics, the inter-grain component has always the same shape, with a minimum height decreasing for growing $\nu$, whereas the intra-grain behavior, better shown in the zoom in Fig. 13, is characterized by a shape modification: in particular, for increasing frequencies a positive peak near $T_c$ decreases, changing its temperature, then it becomes a negative relative minimum and finally merges with the inter-grain contribution.  By taking into account both the frequency and amplitude dependence of the harmonics, we conclude that the vortices moving inside the grains respond in different dynamical way, to the application of the AC field, with respect to those vortices moving across the grains.
Finally, the third harmonics Cole-Cole plots are almost completely in the left semi-plane at the lowest frequency (107 Hz). For increasing frequencies, they tend to the right semi-plane and their area decreases.  The curves at high frequencies correspond to a tendency toward the {\it Bean} curves, which are closed loops all staying in the right half-plane. These behaviors cannot be interpreted in terms of a standard vortex glass phase and are also opposite to the measured YBCO third harmonics Cole-Cole plots\cite{adesso08}, which tend to the left half-plane for increasing frequencies, with an almost constant area.

%
\begin{figure}[!t]
\includegraphics[width=0.42\textwidth]{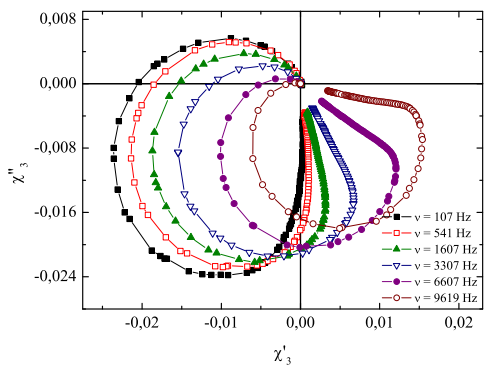}
\caption{ (Color online) Third harmonics Cole-Cole plots at various frequencies of the AC magnetic field at $h_{AC}$=1~Oe.}
\end{figure}
%

\section{Conclusions}

The first and third harmonics of the AC magnetic susceptibility have been measured on a LaOFFeAs granular sample as a function of the temperature, at various amplitudes and frequencies of the AC magnetic field. A comparison between the obtained results and the magnetic behavior of the cuprate superconductors as well as the MgB$_2$ has been performed. The general characteristics, mainly associated with the granular nature of the sample, have been detected in this material as in the previously analyzed superconductors. Nevertheless, some fundamental differences have been revealed between the LaOFFeAs magnetic response and the high-$T_c$, namely YBCO and BSCCO, and MgB$_2$ curves, which can be ascribed to different flux pinning and/or vortex dynamical properties. In particular, an opposite dependence on the amplitude of the AC field in the inter grain component of the first harmonics Cole-Cole plots implies that the standard pinning and flux creep are appropriate enough to analyze the inter-grain AC magnetic response coming from this new material. On the other side, a further amplitude dependence has to be included in the pinning potential to interpret the curves related to the AC magnetic response of the separated grains.
The dissipative phenomena due to the vortex motion have been also investigated in this iron based superconductor. In particular, by the experimental observation of different onset temperatures in the first and third harmonics, the linear TAFF has been detected. Moreover, by a combined analysis of the first and the third harmonics, the non-linear flux creep has been identified in the inter-grain component. The frequency dependence in the imaginary part of the first harmonics, both as a function of the temperature and of the real part of the first harmonics, furnishes some indications about the existence of a vortex glass phase, but they are not enough to identify it univocally. The difficulties to detect a vortex dynamical regime which main influences the full magnetic response of the analyzed superconductor are mainly due to the granular nature of the sample. In particular, the analysis of the amplitude and frequency dependence in the inter-grain and intra-grain components shows that the vortex dynamics occurring inside the grains is different from that produced between the grains. Moreover, a shape change in the imaginary part of the third harmonics in the intra-grain component, both with the amplitude and the frequency of the AC magnetic field, together with a shape modification of the real part of the third harmonics with the frequency, in the inter-grain component, are indicative that the vortex dynamics is changing with the external parameters. This last observation suggests that, in order to reproduce the experimental data, it is not possible to perform numerical simulations by using the same flux creep model in all the curves, but a more complex  analysis is necessary to individuate the better model to be used in correspondence to the different external magnetic solicitations.

\section*{ACKNOWLEDGMENTS}
We kindly thank A. Vecchione and R. Fittipaldi for useful and fruitful discussions.

\end{document}